\documentstyle[11pt]{article}
\begin{document}
\thispagestyle{empty}
\begin{flushright} February 2000\\
\end{flushright}
\vspace{0.5in}
\begin{center}

{\Large \bf Comment on ``Closing the neutrinoless double beta decay window 
into VEP and/or VLI''}

\vspace{1.0in}

{\bf H.V. Klapdor--Kleingrothaus$^1$, H. P\"as$^1$ and U. Sarkar$^2$\\}

\vspace{0.2in}

\noindent {$^1$ \sl Max--Planck--Institut f\"ur Kernphysik,
P.O. Box 103980, D--69029 Heidelberg,
Germany \\}
\noindent{$^2$ \sl Physical Research Laboratory, Ahmedabad 380 009, India\\}

\vspace{1.0in}

\begin{abstract}

The constraints from the non-observation of neutrinoless double beta decay
on the  violations of Lorentz invariance (VLI) or violations of the 
equivalence principle (VEP) have recently been re-examined and it was 
claimed that the constraints are not valid \cite{wrong}.
In this reply we 
point out that this statement is not correct and prove that the arguments 
given are wrong.
\end{abstract}
\end{center}

\newpage
\baselineskip 18pt

In recent times there have been many attempts to find out to what 
accuracy gravitational laws are correct in the neutrino sector.
To constrain the amount of violation of the equivalence principle
(VEP) from neutrino oscillation experiments, one assumes that the
neutrinos of different generations have different characteristic 
couplings to gravity \cite{vep1,vep2}, while to constrain the 
amount of violation of local Lorentz invariance (VLI) one assumes
that neutrinos of different generations have characteristic 
maximum attainable velocities \cite{vli1,vli2}. Some time back 
we pointed out that in both these cases it is possible to constrain
some otherwise unconstrained region in the parameter space from 
neutrinoless double beta decay \cite{beta}. However, a recent paper
\cite{wrong} came to the contrary conclusions.
In this comment we point
out what went wrong in their arguments. 

The decay rate for the neutrinoless double beta decay is given by,
\begin{equation} 
[T_{1/2}^{0\nu\beta\beta}]^{-1}=\frac{M_+^2}{m_e^2} 
G_{01} |ME|^2,
\end{equation} 
where $ME$ denotes the nuclear matrix element, $G_{01}$ corresponds 
to the phase space factor defined in \cite{doi} and $m_e$ is the 
electron mass. The momentum dependence of $M_+$ must be absorbed 
into the nuclear matrix element $|ME|$. 
It will be pointed out in the following, that the dominant contribution
to neutrinoless double beta decay in the case of VLI or VEP results from the 
momentum dependence of the observable itself and has been missed in the 
ansatz in \cite{wrong}.

We shall now present a more detail explanation of this argument.
Reference \cite{wrong} starts with the neutrino propagator 
\begin{equation} 
\int d^4 q \frac{e^{-i q (x-y)} \langle m \rangle c_a^2}{m^2 c_a^4 
- q_0^2 c_a^2 + \vec{q}^2 c_a^2 }
\end{equation} 
with the standard $0\nu\beta\beta$ observable $\langle m \rangle$, the
neutrino four momentum $q$ and the characteristic maximal velocity $c_a$.
They neglect $m$ in the denominator, so that $c_a$ drops
out and the decay rate becomes independent of $c_a$.

However, to derive the double beta decay rate correctly, 
one has to start from the Hamiltonian level. 
In the original paper \cite{beta} it has been shown 
that the propagator (or the $0\nu\beta\beta$ observable) itself is changed 
when the maximum attainable velocities of different neutrino species
are different. Since
\begin{eqnarray} 
H&=&\vec{q} c_a + \frac{m^2 c_a^4}{2 \vec{q} c_a} \nonumber  \\
&=& \vec{q} I + \frac {m^{(*)2} c_a^4}{2 \vec{q} c_a} \label{1}
\end{eqnarray} 
with $c_a = I + \delta v$ and $m^{(*)2}=m^2 + 2 \vec{q^2} c_a \delta v $
an additional contribution to the effective mass 
is obtained $\propto \vec{q^2} \delta v$.
This mass-like term has a $\vec{q^2}$ enhancement and is not
proportional to the small neutrino mass. 
This contribution was included through the nuclear 
matrix element $|ME|$. But in reference \cite{wrong} the
authors started with a propagator in the zero momentum transfer 
approximation and obviously did not get this additional important 
term. If the observable $M_+$ is assumed to be momentum-independent in the 
following (i.e. neglecting nuclear recoil), of course the momentum dependence 
has to be included in the nuclear matrix element. 

In the original paper \cite{beta}, a bound on VEP or VLI was 
presented in the small mixing region assuming conservatively 
$\langle m \rangle \simeq 0$ and $\delta{m} \leq \bar{m}$.
Due to the $q^2$ enhancement the nuclear matrix elements of the 
mass mechanism were replaced by $\frac{m_p}{R}\cdot 
(M_F^{'}-M_{GT}^{'})$ with the nuclear radius $R$ and the 
proton mass $m_p$, 
which have been calculated in \cite{mat}.
Inserting the recent half life limit obtained from the 
Heidelberg--Moscow experiment \cite{double},  
a bound on the amount of tensorial gravitational interactions 
as a function of the average neutrino mass $\bar{m}$ was 
presented \cite{beta}. {\bf It is obvious that
if one ignores the momentum enhancement of the nuclear matrix
element (as it was done in ref \cite{wrong}), the neutrinoless
double beta decay cannot give any significant constraint}. In fact,
including the momentum enhancement of the nuclear matrix elements
not only these constraints are significant, they will further
improve by 1--2 orders of magnitude with the GENIUS proposal of the 
Heidelberg group \cite{gen}. 

{\bf Acknowledgement}
We thank G. Bhattacharyya for useful discussions.

\end{document}